\documentclass[%
reprint,
amsmath,amssymb,
aps,
]{revtex4-2}
\usepackage{amsmath}
\usepackage{booktabs}
\usepackage{graphicx}
\usepackage{dcolumn}
\usepackage{subfigure}
\usepackage{microtype}
\usepackage{bm}
\usepackage{hyperref}
\usepackage{xcolor}

\begin{document}

\preprint{APS/123-QED}
\title{{Dynamical backreaction of a mass-acquiring scalar field on first-order phase transitions}}

\author{Yuan-Jie Li}
\email{liyuanjie23@mails.ucas.ac.cn}
\affiliation{%
International Centre for Theoretical Physics Asia-Pacific, University of Chinese Academy
of Sciences, 100190 Beijing, China
}%
\affiliation{%
Taiji Laboratory for Gravitational Wave Universe, University of Chinese Academy of Sciences, 100049 Beijing, China
}%
\author{Jing Liu}
\email{liujing@ucas.ac.cn}
\affiliation{%
International Centre for Theoretical Physics Asia-Pacific, University of Chinese Academy
of Sciences, 100190 Beijing, China
}%
\affiliation{%
Taiji Laboratory for Gravitational Wave Universe, University of Chinese Academy of Sciences, 100049 Beijing, China
}%
\author{Zong-Kuan Guo}
\email{guozk@itp.ac.cn}
\affiliation{%
Institute of Theoretical Physics, Chinese
Academy of Sciences, P.O. Box 2735, Beijing 100190, China
}%
\affiliation{%
School of Physical Sciences, University of Chinese Academy of Sciences, No.19A Yuquan
Road, Beijing 100049, China
}%
\affiliation{%
School of Fundamental Physics and Mathematical Sciences, Hangzhou Institute for Advanced Study, University of Chinese Academy of Sciences, Hangzhou 310024, China
}%
\begin{abstract}
    Phase transitions in the early Universe give rise to effective masses for massless fields in the symmetry-broken phase. {We perform lattice simulations to study the dynamical impact of a mass-acquiring spectator field on the evolution of first-order phase transitions and the associated gravitational-wave production, while keeping the effective potential responsible for bubble nucleation fixed.}
    In addition to the well-known friction effects, we identify a novel effect that significantly enhances the strength of first-order phase transitions.
    {In contrast to the general scenario, although the effective potential governs the tunneling rate, the amplitude of the $\chi$ field is strongly suppressed inside the true vacuum bubble, resulting in a faster bubble expansion than predicted by the effective potential alone.}
    The amplitude of the mass-acquiring field is highly suppressed in the true vacuum bubbles, resulting in additional release of vacuum energy that concentrate on the bubble walls. {We further develop an analytical framework that not only explains our numerical results but can also be used to improve the estimation of gravitational-wave signals in related phase-transition scenarios.}
    

\end{abstract}

\maketitle

\section{Introduction}

First-order phase transitions~(FOPTs) in the early Universe are initiated by the nucleation of true vacuum bubbles through quantum tunneling~\cite{Coleman:1977py,Callan:1977pt,Coleman:1980aw,Linde:1981zj,Kusenko:1997hj,Konoplich:1999qq,Khlopov:2000js}. 
The subsequent release of vacuum energy accelerates the bubble walls and result in the generation of gravitational waves (GWs) through bubble collisions, sound waves, and turbulence~\cite{Kosowsky:1991ua,Kosowsky:1992vn, Huber:2008hg, Hindmarsh:2013xza, Mazumdar:2018dfl, Bian:2025ifp,Linde:1981zj,Caprini:2015zlo}. These transitions represent one of the most promising cosmological origins of stochastic GW backgrounds, potentially observable by upcoming space-based interferometers including LISA~\cite{Kahniashvili:2008pf, Caprini:2019egz, Schmitz:2020rag, Boileau:2022ter}, Taiji~\cite{Ren:2023yec, Huang:2025uer}, and TianQin~\cite{Cheng:2022vct, Li:2024rnk}.

Accurate predictions of the GW energy spectrum require detailed modeling of the underlying dynamics, including the bubble wall acceleration, the energy transfer processes, and the post-collision field evolution. Multiple analytical and numerical techniques have been developed, including the envelope approximation~\cite{Kosowsky:1992vn, Caprini:2007xq}, hydrodynamic treatments~\cite{Hindmarsh:2013xza, Konstandin:2017sat,Tian:2024ysd}, and fully field-theoretic lattice simulations~\cite{Hindmarsh:2015qta,Cutting:2018tjt, Ellis:2019oqb,Di:2020kbw,Zhao:2022cnn,Li:2023yzq,Zou:2025sow}. These studies have systematically characterized how key physical parameters, particularly the wall thickness, the wall velocity, and the out-of-equilibrium scalar field evolution, determine the spectral features of the resulting GW signals.

More recently, increasing attention has turned to the scenarios where the scalar field that triggers the FOPT couples to additional fields which acquire effective masses in the true vacuum. Such mass-generating mechanism has been incorporated into a variety of beyond Standard Model scenarios, including filtered dark matter~\cite{Baker:2019ndr,Chway:2019kft,Ahmadvand:2021vxs,Jiang:2023nkj}, Fermi-ball dark matter~\cite{Hong:2020est,Kawana:2021tde,Flores:2023zpf}, non-thermal particle production in the dark sector~\cite{Reece:2015lch,Giudice:2024tcp}, highly interactive particle relics dark matter~\cite{Elor:2021swj}, and the Q-ball formation~\cite{Krylov:2013qe,Jiang:2024zrb}.  The presence of mass-acquiring fields also alters the effective potential of FOPTs and lead to energy redistribution. In such cases,  even if the mass-acquiring fields are energetically subdominant, they may alter the FOPT dynamics and modify the GW signatures~\cite{Espinosa:2008kw}. 


In this work, we numerically investigate the impact of a mass-acquiring scalar field during an FOPT, focusing on its effect on the bubble wall evolution and the resulting GW energy spectrum. {Throughout this paper the phase transition is driven solely by the scalar field $\phi$. Both the false and true vacua satisfy $\langle \chi \rangle = 0$, so the transition proceeds as
\begin{equation*}
(\phi,\chi):(0,0)\rightarrow(\phi_\mathrm{b},0),
\end{equation*}
and the $O(4)$-symmetric bounce trajectory is confined to the $\chi=0$ axis. Our setup therefore does not realise a two-field phase transition with non-trivial tunnelling in the $(\phi,\chi)$ plane; instead, $\chi$ plays the role of a spectator field.}

We fix the effective potential for the first-order phase transition (FOPT) that determines the bubble nucleation rate, and numerically compare results across different values of both the coupling constant and the mass-acquiring field fraction. {This setup allows us to isolate the dynamical effects induced by the field evolution from the trivial modifications of the potential shape, ensuring that the observed differences do not originate from a change in the effective potential itself.} By analyzing simulation results, it was found that the amplitude of the mass-acquiring field within the true vacuum bubbles significantly decreases, which consequently reducing the minimum of the true vacuum potential. This effect is equivalent to an effective enhancement of the transition strength, leading to an increase in bubble wall velocity and an amplification of GWs. We also analyze the redistribution of the released vacuum energy among different components and develop an analytical power-law expression that explains the observed GW amplitude. 

This paper is organized as follows. In Section~\ref{sec:model}, we present the theoretical setup of the FOPT, including the effective potential, the equations of motion, and the bubble nucleation conditions. Section~\ref{sec:methods} describes the lattice simulation framework and the initial conditions. In In Section~\ref{sec:evolution} and Section~\ref{sec:gw}, we respectively analyze the FOPT dynamics and the GW signatures based on the numerical results, focusing on the impact of the mass-acquiring field. Finally, we summarize our findings and discuss their broader implications in Section~\ref{sec:summary}. For convenience, we apply $c=\hbar=1$ throughout this paper.

\section{the Model}\label{sec:model}
\subsection{The Effective Potential}
The effective potential for the scalar fields $\phi$ and $\chi$ is defined as
\begin{align}\label{eq:full_effective_potential}
    V(\phi, \chi) = V(\phi) + V_{\phi\chi}(\phi, \chi)\,,
\end{align}
where 
\begin{align}
V(\phi) \equiv \tfrac{1}{4} \lambda_\phi \phi^4 - \tfrac{\beta}{3} \phi^3 + \tfrac{1}{2} \mu_\phi^2 \phi^2
\end{align}
is the uncoupled effective potential of $\phi$, which possesses two local minima and thus allows for quantum tunneling. The field $\phi$ couples to another massless scalar field $\chi$ through the interaction term $V_{\phi\chi}(\phi, \chi) \equiv \lambda_{\phi\chi} \phi^2 \chi^2$. We focus on the case that FOPTs occur on timescales much shorter than the Hubble time at the transition epoch \( H_*^{-1} \), such that the expansion of the Universe can be neglected. The equations of motion for the scalar fields are given by
\begin{align}\label{eq:EoM}
    \Box \phi + \frac{\partial V(\phi, \chi)}{\partial \phi} = 0, \quad
    \Box \chi + \frac{\partial V(\phi, \chi)}{\partial \chi} = 0.
\end{align}

{Because the interaction term does not induce a vacuum expectation value for $\chi$, the relevant minima of $V(\phi,\chi)$ for the phase transition are located at $(\phi,\chi)=(0,0)$ and $(\phi_\mathrm{b},0)$, with $\langle\chi\rangle = 0$ in both phases. Any $O(4)$-symmetric bounce solution interpolating between these vacua therefore lies entirely on the $\chi = 0$ axis. Solving the tunnelling problem with the full two-field potential $V(\phi,\chi)$ is then equivalent to using the one-dimensional potential $V(\phi)$: the field $\chi$ is effectively kept frozen, the Euclidean action and the critical bubble profile coincide with the standard single-field result, and the effect of $\chi$ fluctuations on the tunnelling action and on the nucleation conditions is not captured. For this reason we do not use $V(\phi,\chi)$ or $V(\phi)$ to compute the bounce, but instead adopt the reduced potential $V_{\mathrm{re}}(\phi)$ introduced below, in which the leading impact of $\chi$ fluctuations enters through $\langle\delta\chi^2\rangle$.}

The effect of the interaction term comes in two aspects.
On one hand, as the field \( \phi \) acquires a non-zero vacuum expectation value \( \phi_{\text{b}} \) in the true vacuum, \( \chi \) acquires an effective mass through the interaction term 
\begin{align}
    m_\chi^2 = 2 \lambda_{\phi\chi} \phi_{\text{b}}^2.
\end{align}

On the other hand, the existence of $\chi$ also alters the form of the FOPT potential and changes the prediction of the bubble nucleation rate. Since the wavelength of $\chi$ is in general much smaller than the vacuum bubbles, we can apply the mean-field approximation to obtain the reduced effective potential
\begin{align}\label{eq:Vre}
V_{\text{re}}(\phi) = \frac{1}{4} \lambda_\phi \phi^4 - \frac{\beta}{3} \phi^3 + \frac{1}{2} M_\phi^2 \phi^2\,, 
\end{align}
where \( M_\phi^2 \equiv \mu_\phi^2 + 2\lambda_{\phi\chi} \langle \delta\chi^2 \rangle \) is the effective mass of \( \phi \) and the angle brackets presents the spatial average. We use the symbol $\delta\chi$ since $\langle\chi\rangle$ vanishes. Note that Eq.~\eqref{eq:Vre} is only responsible for the bubble nucleation, while the numerical calculations of field evolutions are governed by Eq.~\eqref{eq:EoM}. {It is worth noting that this analysis is applicable to the weak-coupling regime, where the interaction strength $\lambda_{\phi\chi}$ is small. In this limit, loop effects are subdominant and can be safely neglected within the perturbative regime considered in this work~\cite{Coleman:1973jx,Coleman:1977py,Quiros:1994dr}.}

For a fast FOPT, the energy scale of the Universe remains roughly unchanged, so  quantum tunneling only occurs when $V_{\mathrm{re}}$, or equivalently $M_{\phi}$, reaches the threshold. The parameter $\mu_{\phi}$ generally decreases as the temperature drops. 
Therefore, the presence of $2\lambda_{\phi\chi} \langle \delta\chi^2 \rangle$ does not change the effective potential $V_{\mathrm{re}}$ at the the onset of the FOPT, it is the threshold that determines the beginning of sufficient bubble nucleation. For larger $2\lambda_{\phi\chi} \langle \delta\chi^2 \rangle$, the FOPT is postponed slightly to render a smaller $\mu_{\phi}$, leaving $M_{\phi}$ almost unchanged, as the threshold determines. Consequently, it is reasonable to fix $M_{\phi}$ as a constant among different conditions of $\lambda_{\phi\chi}$ and the amplitude of $\chi$. Furthermore, the direct influence of the interaction terms on the tunneling rate can also be excluded. {In other words, the effective potential responsible for bubble nucleation is kept unchanged when introducing the additional scalar field $\chi$, so that the subsequent differences in the evolution of fields and the GW signal originate purely from the dynamical evolution of the coupled fields rather than from a static modification of the potential. As we will see, as the evolution proceeds, the expansion of the bubble wall is not dictated by $V_{\mathrm{re}}(\phi)$, but instead aligns more closely with the prediction of $V(\phi)$.}

\begin{figure}[htb]
    \includegraphics[width=\linewidth]{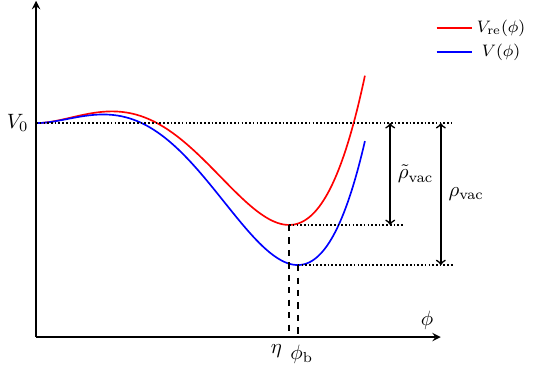}
    \caption{Schematic comparison between the effective potential without the interaction term $V(\phi)$ (blue) and the reduced potential \( V_{\text{re}}(\phi) \) (red), where the latter is fixed in our subsequent analysis.
The potential energy density at the false vacuum is labeled \( V_0 \) and the vacuum expectation values in the broken phase are denoted \( \phi_{\text{b}} \) and \( \eta \) for \( V(\phi) \) and \( V_{\text{re}}(\phi) \), respectively. 
The corresponding vacuum energy differences are labeled \( \rho_{\text{vac}} \) and \( \tilde{\rho}_{\text{vac}} \), respectively.
}
    \label{fig:potential}
\end{figure}

To illustrate the impact of $\chi$ on \( V_{\text{re}}(\phi) \), we compare $V(\phi, \chi)$ with \( V_{\text{re}}(\phi) \) in Fig.~\ref{fig:potential}, where the vacuum expectation values are \( \eta \) and \( \phi_{\text{b}} \) for the two potentials, respectively. The potential energy density at the false vacuum is labeled \( V_0 \), and the corresponding energy density differences between the true and false vacua are labeled $\tilde{\rho}_{\mathrm{vac}}$ and $V(\phi)$, respectively.

\subsection{Bubble Nucleation}
In an FOPT, bubble nucleation proceeds via quantum tunneling through the potential barrier. We consider the zero-temperature case, where the solution of the initial bubble satisfies the \( O(4) \) symmetry~\cite{Coleman:1977py,Linde:1980tt,Linde:1981zj,Salvio:2016mvj}. As discussed in the last section, the tunneling process is governed by the reduced potential \( V_{\text{re}} \). The corresponding Euclidean action is given by
\begin{align}\label{eq:profile}
        S_4 = 2\pi^2 \int_0^{\infty} d\rho\, \rho^3 \left[ \frac{1}{2} \left( \frac{d\phi}{d\rho} \right)^2 + V_{\text{re}}(\phi) \right],
\end{align}
where \( \rho = \sqrt{\tau^2 + \mathbf{r}^2} \) is the radial coordinate in four-dimensional Euclidean spacetime, and \( \tau = \mathrm{i}t \) denotes the Euclidean time.

When the radius of a critical bubble is much larger than the thickness of the bubble wall, i.e., in the thin-wall limit, the scalar field profile of the critical bubble is given by
\begin{align}\label{eq:profile}
    \phi_{\text{c}}(r) = \frac{\eta}{2} \left[ 1 - \tanh\left( \frac{r - R_\text{c}^{\text{tw}}}{l_0^{\text{tw}}} \right) \right],
\end{align}
where \( R_{\text{c}}^{\text{tw}} \) and \( l_0^{\text{tw}} \) denote the radius of the critical bubble and the thickness of the critical bubble wall, respectively.

Depending on the time dependence of the Euclidean action \( S_4(t) \), the bubble nucleation during a FOPT can be categorized into three typical modes. In the exponential nucleation regime, where \( S_4(t) \) decreases slowly with time, the nucleation rate grows approximately as \( p(t) \sim e^{\beta t} \)~\cite{Enqvist:1991xw}. If \( S_4(t) \) remains nearly constant, the nucleation rate is approximately time-independent, i.e., \( p(t) = \text{const} \), which defines the constant nucleation regime~\cite{GarciaGarcia:2016xgv}. When \( S_4(t) \) has a minimum at some time \( t_0 \), nucleation becomes sharply peaked, corresponding to simultaneous nucleation, where most bubbles nucleate nearly simultaneously around \( t_0 \)~\cite{Jinno:2017ixd}. Since the GW signal depends primarily on the bubble number density rather than the detailed nucleation history~\cite{Cutting:2018tjt,Cutting:2020nla}, we adopt the simultaneous nucleation scenario throughout this work.

\section{Numerical Methods}\label{sec:methods}
In this section, we present our numerical method to simulate the field evolutions and the GW production.

We numerically solve the equations of motion Eq.~\eqref{eq:EoM} in Minkowski spacetime on a three-dimensional cubic lattice with periodic boundary conditions. We use the open-source C++ package \textit{\texttt{CosmoLattice}} (version 1.2) to conduct the simulations~\cite{Figueroa:2020rrl,Figueroa:2021yhd}. Spatial derivatives are computed using a second-order central finite difference method, and the time evolution is advanced using the leapfrog algorithm.

Each simulation is performed on a cubic lattice with $N = 512$ grid points per spatial dimension, corresponding to a total volume $\mathcal{V} = (N \Delta x)^3$. The box size is set to be $L \equiv N\Delta x = 128\,\omega_*^{-1}$, where \( \omega_* \equiv \sqrt{\lambda_\phi} \eta =(4\Delta x)^{-1}\) is a characteristic mass scale. The time step is chosen as $\Delta t = 0.2 \Delta x$ to ensure numerical stability.

We obtain the critical bubble profile using \textit{\texttt{FindBounce}}, a package that numerically solves the bounce equations {in the reduced potential $V_{\mathrm{re}}(\phi)$} via the shooting method~\cite{Guada:2020xnz}. The resulting profile is well fitted by the form of Eq.~\eqref{eq:profile}, which corresponds to the analytical solution in the thin-wall approximation. The parameters are set to be \( V_0 = 0.2\, \omega_*^2 \eta^2 \), \( \beta = 1.3\, \omega_*^2 \eta^{-1} \), and \( M_\phi^2 = 0.3\, \omega_*^2 \). The corresponding critical bubble radius and wall thickness are \( R_c = 10.06\, \omega_*^{-1} \) and \( l_0 = 3.05\, \omega_*^{-1} \), respectively.
At the initial time $t_i = 0$, $N_b$ critical bubbles are nucleated to trigger the FOPT. The centers of any two bubbles are separated by more than \( 2R_c \) to avoid initial overlap. 
However, Equation~\eqref{eq:profile} indicates that the bubble profile exhibits a tail extending to infinity. To self-consistently place multiple bubbles without introducing sharp cutoffs, the bubbles are initialized by superimposing the profiles as~\cite{Cutting:2018tjt}
\begin{align}
    \phi(\mathbf{x}) = \sqrt{ \sum_{i=1}^{N} \phi_{\text{c}}^2(|\mathbf{x} - \mathbf{x}_i|) },
\end{align}
where \( \mathbf{x}_i \) denotes the center of the \( i \)-th bubble.

In multi-bubble simulations, we take $N_b = 32$, yielding an average bubble separation
\begin{align}
    R_* \equiv \left( \frac{\mathcal{V}}{N_b} \right)^{1/3} = 39.37\,\omega_*^{-1},
\end{align}
which exceeds the critical radius $R_c$ and thereby ensures successful bubble initialization. Proper resolution of the bubble wall dynamics requires that the Lorentz-contracted wall thickness, $l_* \equiv l_0 / \gamma_*$, satisfies $l_* \gg \Delta x$, where $\gamma_*$ is the expected Lorentz factor of the bubble front at collision. As shown in the Appendix~\ref{app:lattice_effect}, for our parameter choices guarantees $\gamma_* \leq 4$, lattice discretization effects remain negligible.


To investigate the impact of $\chi$ and the interaction term, we take different values of the coupling constant  \( \lambda_{\phi\chi} \) and fluctuation amplitude \( A \) in our simulations. The parameter $\mu_{\phi}$ is changed correspondingly to ensure $M_{\phi}$ and $V_{\mathrm{re}}$ unchanged.
The simulation parameters are listed in Table~\ref{tab:parameters}.

The initial condition for \( \chi \) is set independently via quantum vacuum fluctuations, characterized by the power spectrum
\begin{align}
    \mathcal{P}_{\delta\chi}(k) = \frac{A^2}{k},
\end{align}
where \( A \) is a free parameter denoting the fluctuation amplitude. This leads to a variance
\begin{align}
    \langle \delta\chi^2 \rangle = \int d\log k\, \frac{k^3}{2\pi^2} \mathcal{P}_{\delta\chi}(k) 
    \simeq \frac{A^2 k_{\text{max}}^2}{8\pi^2},
\end{align}
with \( k_{\text{max}} \) the ultraviolet cutoff. In each simulation, we fix \( k_{\text{max}}=\eta \) as a constant. As a result, the effective mass in the reduced potential \( V_{\text{re}} \) becomes
\begin{align}
    M_\phi^2 = \mu_\phi^2 + \frac{\lambda_{\phi\chi} A^2 k_{\text{max}}^2}{4\pi^2}.
\end{align}

\begin{table*}[htbp]
	\centering
	\begin{tabular}{c*{12}{|c}} 
		\hline
        \hline
        $\lambda_{\phi\chi}/\lambda_{\phi}$ 
		& 0 & 0.5 & 1 & 2 & 5 & 8 & 10 & 1 & 1 & 1 & 1 & 1 \\
		\hline
		$A$ 
		& -- & 1 & 1 & 1 & 1 & 1 & 1 & 1.5 & 2 & 2.5 & 3 & 4 \\
		\hline
		$\mu_{\phi}^{2}/\omega_{*}^{2}$ 
		& 0.3 & 0.287335 & 0.274670 & 0.249339 & 0.173349 & 0.0973576 
		& 0.046697 & 0.243007 & 0.198679 & 0.141686 & 0.0720273 & -0.105285 \\
        \hline
		\hline
	\end{tabular}
    \caption{The parameter sets used in the simulations. The coupling constant \( \lambda_{\phi\chi} \) and the fluctuation amplitude \( A \) are changed in different simulations, and \( \mu_\phi^2 \) is also adjusted to keep the reduced potential \( V_{\text{re}}(\phi) \) fixed.}
    \label{tab:parameters}
\end{table*}

\section{The Evolution}\label{sec:evolution}
\subsection{The Growth of a Single Bubble}\label{sec:growth_of_single_bubble}

\begin{figure*}[htbp]      
    \centering
    \begin{minipage}[htbp]{0.48\textwidth}
        \centering
        \includegraphics[width=\textwidth]{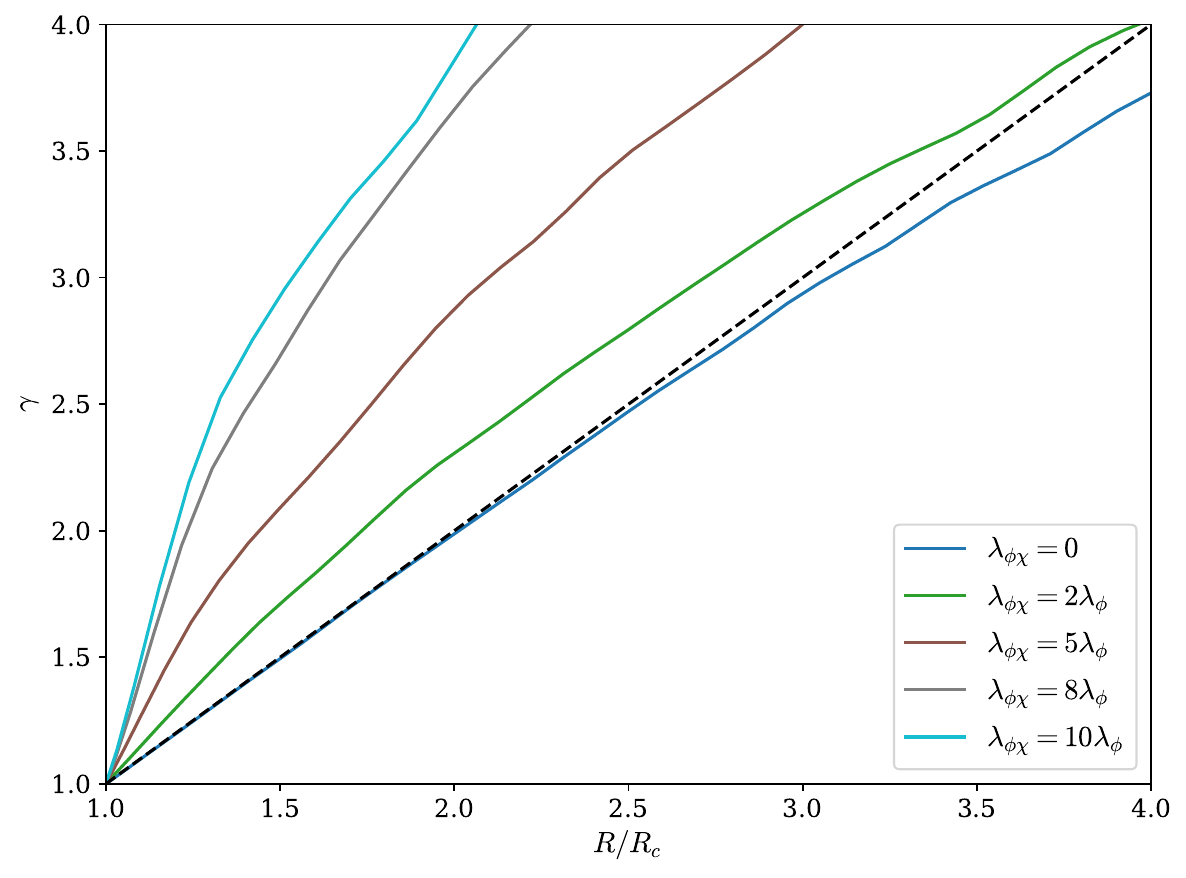}
        (a) Varying \( \lambda_{\phi\chi}/\lambda_\phi \) with fixed \( A = 1 \).
    \end{minipage}
    \begin{minipage}[htbp]{0.48\textwidth}
        \centering
        \includegraphics[width=\textwidth]{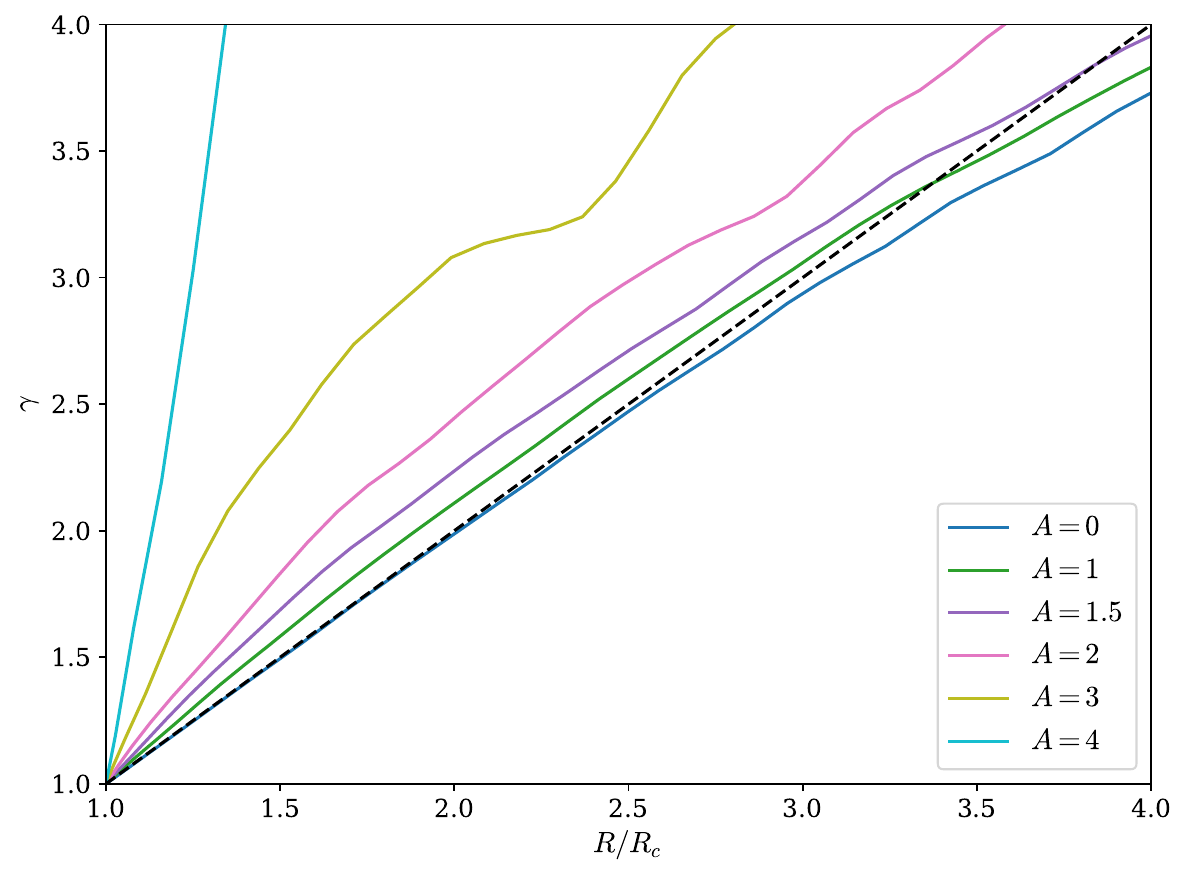}
        (b) Varying \( A \) with fixed \( \lambda_{\phi\chi} = \lambda_\phi \).
    \end{minipage}
    \caption{Lorentz factor \( \gamma \) of bubble walls as a function of normalized radius \( R/R_c \). In both panels, the black dashed line represents the analytical prediction \( \gamma = R/R_c \) from the thin-wall approximation in the absence of coupling. Note that the parameters of light blue curve in the right panel does not appear in the multi-bubble simulations.}
    \label{fig:gamma_vs_R}
\end{figure*}
We start with the expansion of a single bubble to clearly assess the effect of the interaction term on the bubble dynamics, which also verifies that the numerical errors in the simulations remain under control. 
We initialize a single bubble in the lattice and track its evolution. The bubble radius \( R \) and corresponding Lorentz factor \( \gamma \) are determined by counting lattice points where \( \phi > \phi_{\text{b}}/2 \), allowing us to characterize the \( \gamma(R) \) relationship for various values of \( \lambda_{\phi\chi} \) and \( A \).

The relation between the Lorentz factor \( \gamma \) and the normalized radius \( R/R_c \) is illustrated in Fig.~\ref{fig:gamma_vs_R}. The black dashed line in both panels shows the analytical prediction from energy conservation in the thin-wall approximation without the interaction term~\cite{Coleman:1977py,Linde:1981zj}, 
\begin{align}\label{eq:gamma_r}
    \gamma = R/R_c.
\end{align}

We observe that increasing either \( \lambda_{\phi\chi} \) or \( A \) leads to faster acceleration of the bubble wall. These results suggest that, at fixed \( V_{\text{re}}(\phi) \), increasing the coupling constant or the fluctuation amplitude enhances the effective vacuum energy release, thereby accelerating bubble expansion. The explanation of this phenomenon is provided in detail in the next subsection.

The accelerated expansion of the bubble wall introduces a numerical challenge: as the Lorentz factor $\gamma$ increases, the wall thickness contracts progressively toward the critical spatial resolution limit. This effect places a constraint on our parameter selection. For the $\lambda_{\phi\chi}=0$ and $A=0$ case~(which is referred to as the uncoupled case), the analytical prediction (black dashed line) agrees well with the numerical result (deep blue solid line) for $\gamma\lesssim4$. To ensure numerical precision, we therefore require  the Lorentz factor $\gamma_* \equiv \gamma(R_*/2R_c)$ remain below 4 in multi-bubble simulations. 
For our chosen parameters, namely $R_* = 39.37\,\omega_*^{-1}$ and $R_c = 10.06\,\omega_*^{-1}$, we find $R_*/2R_c \approx 2$. Therefore, we choose the parameter sets in multi-bubble simulations that satisfy the condition $\gamma(2) < 4$.

\subsection{The Multi-Bubble Evolution}
This section analyzes the simulation results to determine the impact of $\chi$ and the interaction term on system dynamics. We also provides analytical interpretations based on these numerical results.

\begin{figure*}[htbp]
    \centering
    \includegraphics[width=\linewidth]{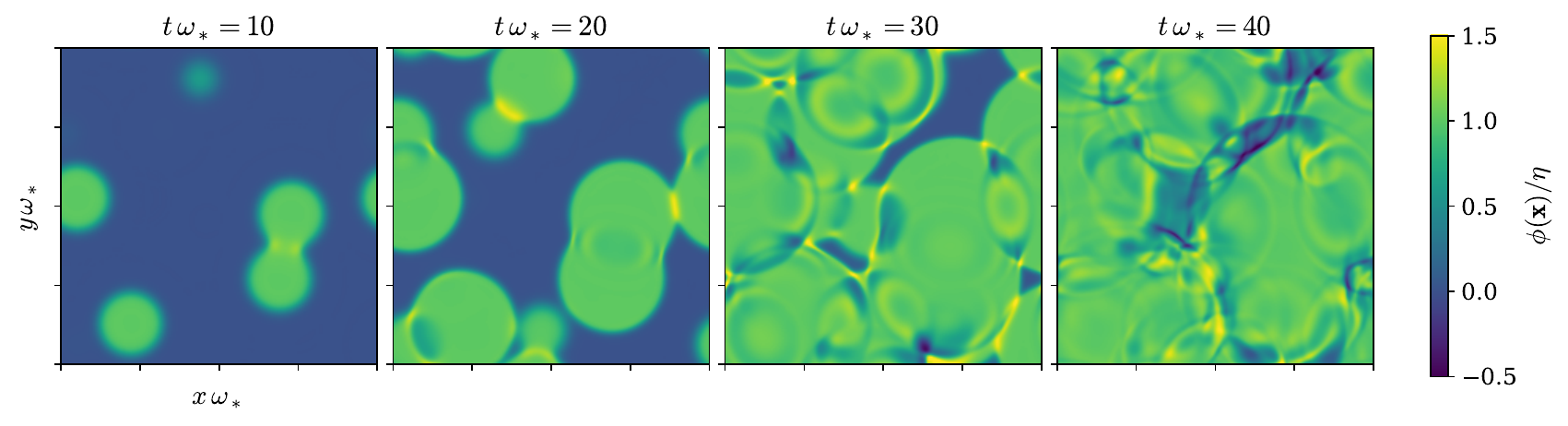}
    \caption{The evolution of the scalar field \(\phi(x)/\eta\) in the absence of coupling (\( \lambda_{\phi\chi} = 0 \) and $A=0$). From left to right, each panel corresponds to snapshots at \( t\omega_* = 10 \), 20, 30, and 40, respectively.}
    \label{fig:phi_evolution_uncoupled}
\end{figure*}

Fig.~\ref{fig:phi_evolution_uncoupled} illustrates the FOPT process without any contribution of $\chi$, i.e., the uncoupled case, $A=0$ and $\lambda_{\phi\chi} = 0$. From left to right, the panels depict the two-dimensional snapshots of $\phi(\mathbf{x})/\eta$ at \( t\omega_* = 10 \), 20, 30, and 40, respectively.
These slices are taken at the central plane $z = L/2$ and capture the dynamics of bubbles. One can clearly see bubble nucleation, expansion, and collision, followed by the completion of the transition as the true vacuum fills the entire space.  After $t\omega_* = 40$, the FOPT is essentially complete, and $\phi$ exhibits oscillations around the true vacuum. 

\begin{figure*}[htbp]
    \centering
    \includegraphics[width=\linewidth]{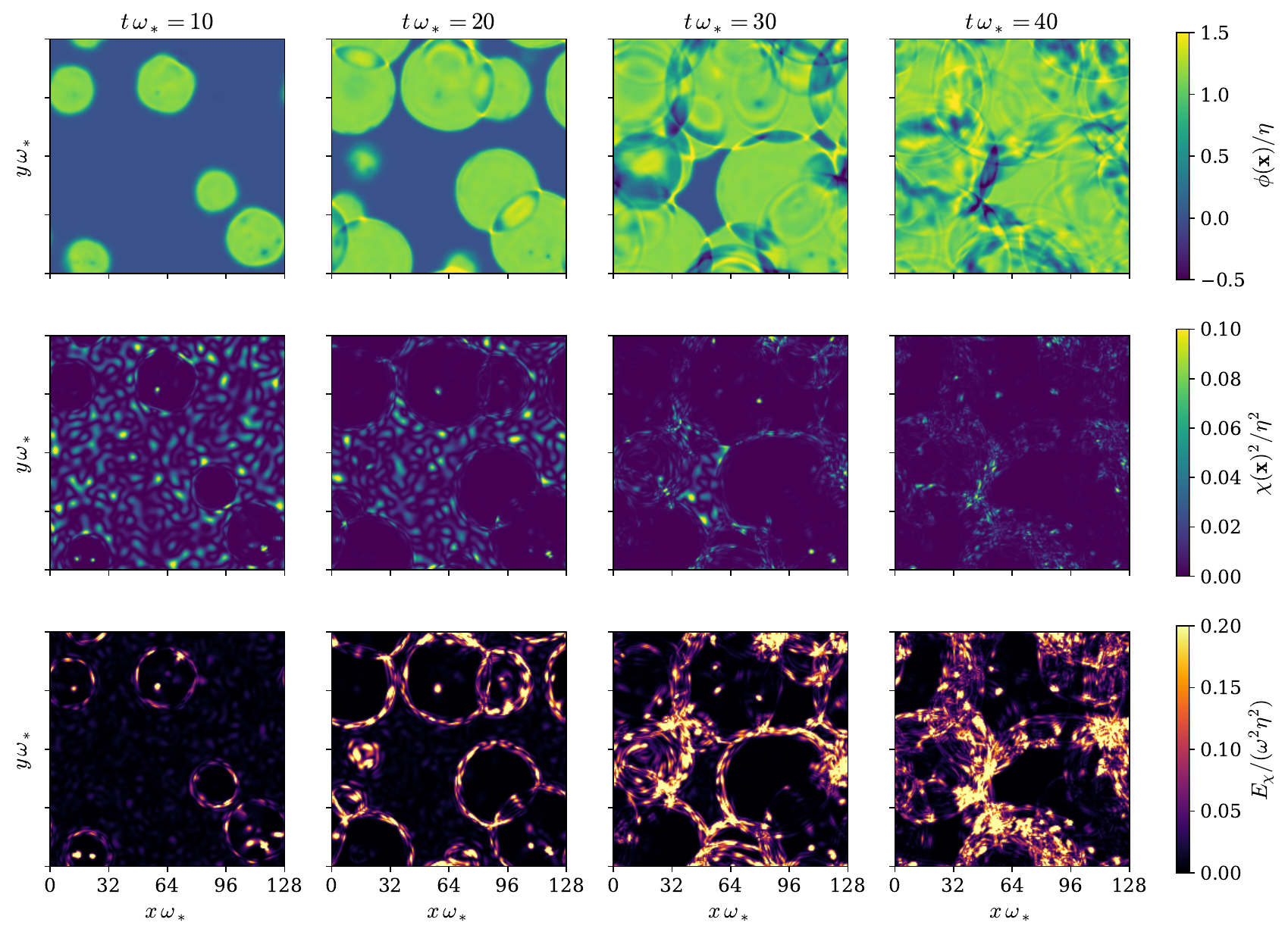}
    \caption{The evolution of $\phi$ and $\chi$ and the energy distribution with \( \lambda_{\phi\chi} = 5\lambda_\phi \) and \( A = 1 \). The top row shows the field configuration of \( \phi(\mathbf{x})/\eta \), the middle row displays \( \chi^2(x,y) / \eta^2 \), and the bottom row presents the total energy density of \( \chi \), namely \( E_\chi = V_{\phi\chi} + K_\chi + G_\chi \), which is normalized by \( \omega_*^2 \eta^2 \). From left to right, each column corresponds to snapshots at \( t\omega_* = 10 \), 20, 30, and 40, respectively.}
    \label{fig:merged_fields_energy_slices}
\end{figure*}

We then focus on how $\phi$ and $\chi$ evolve in the coupled case by considering the representative parameters $\lambda_{\phi\chi} = 5\lambda_{\phi}$ and $A = 1$. To better understand the field dynamics and energy transfer during the phase transition, in Fig.~\ref{fig:merged_fields_energy_slices}, we present not only slices of $\phi$ but also the evolution of $\chi^2$ and the energy density of $\chi$, defined as $E_{\chi} = V_{\phi\chi} + K_{\chi} + G_{\chi}$. Here, $V_{\phi\chi}$, $K_{\chi}$, and $G_{\chi}$ denote the interaction potential energy between $\phi$ and $\chi$, the kinetic energy density of $\chi$, and its gradient energy density, respectively.

In the cases with interactions, the bubbles expand more rapidly, as discussed in Sec.~\ref{sec:growth_of_single_bubble}, leading to an even earlier completion of the FOPT. Moreover, the GW production from bubble collisions does not significantly grow further afterward~\cite{Kosowsky:1992vn,Huber:2008hg,Konstandin:2017sat,Cutting:2018tjt}.

From the medium row of Fig.~\ref{fig:merged_fields_energy_slices}, we can observe that in the true vacuum bubbles, the amplitude of $\chi^2$ rapidly diminish and become negligible. The intuitive explanation is as follows.
For a given Fourier mode $\chi_{k}$, the energy spectrum is given by \( \frac{1}{2} k^{2} \chi_{k}^2 \) for a massless field. If $\chi$ acquires a large effective mass \( m_{\chi} \) with \( m_{\chi} \gg k\), $\chi$'s energy spectrum becomes \( \frac{1}{2} m_{\chi}^{2} \chi_{k}^2 \) since the gradient energy is negligible compared to its potential energy. For the same energy density of $\chi_{k}$, the amplitude \(\delta\phi_{k}\) must then be reduced by a large factor in the true vacuum.
In turn, this effect suppresses $\chi$'s contribution to the potential \( V(\phi, \chi) \) in the true vacuum. Consequently, the potential function in the true vacuum drops from $V_{\mathrm{re}}$ to \( V(\phi) \), shown as the red and blue lines in Fig.~\ref{fig:potential}. This effect effectively increases the vacuum energy released during the FOPT. It also explains the observed enhancement in bubble expansion speed in Sec.~\ref{sec:growth_of_single_bubble} when the interaction term is present.

From the bottom row~(the energy density of $\chi$) we can see that during bubble expansion, part of the energy is transferred to $\chi$ via interactions at the bubble walls, forming a shell-like structure reminiscent of acoustic shells. Moreover, the $\chi$ particles are reflected away by the expanding bubbles, leading to a suppressed energy density of the $\chi$ in the true vacuum. At the end of bubble collisions, the `energy shells' of $\chi$ are finally absorbed by the true vacuum, transforming into oscillations of $\chi$.

We observe some isolated points with large fluctuations, where $\phi$ temporarily remains trapped in the false vacuum due to its interaction with $\chi$. These false vacuum regions is metastable and eventually decay into the true vacuum. These sparse localized objects does not affect our overall analysis and the predictions of GWs.

\begin{figure}[htbp]
    \centering
    \includegraphics[width=\linewidth]{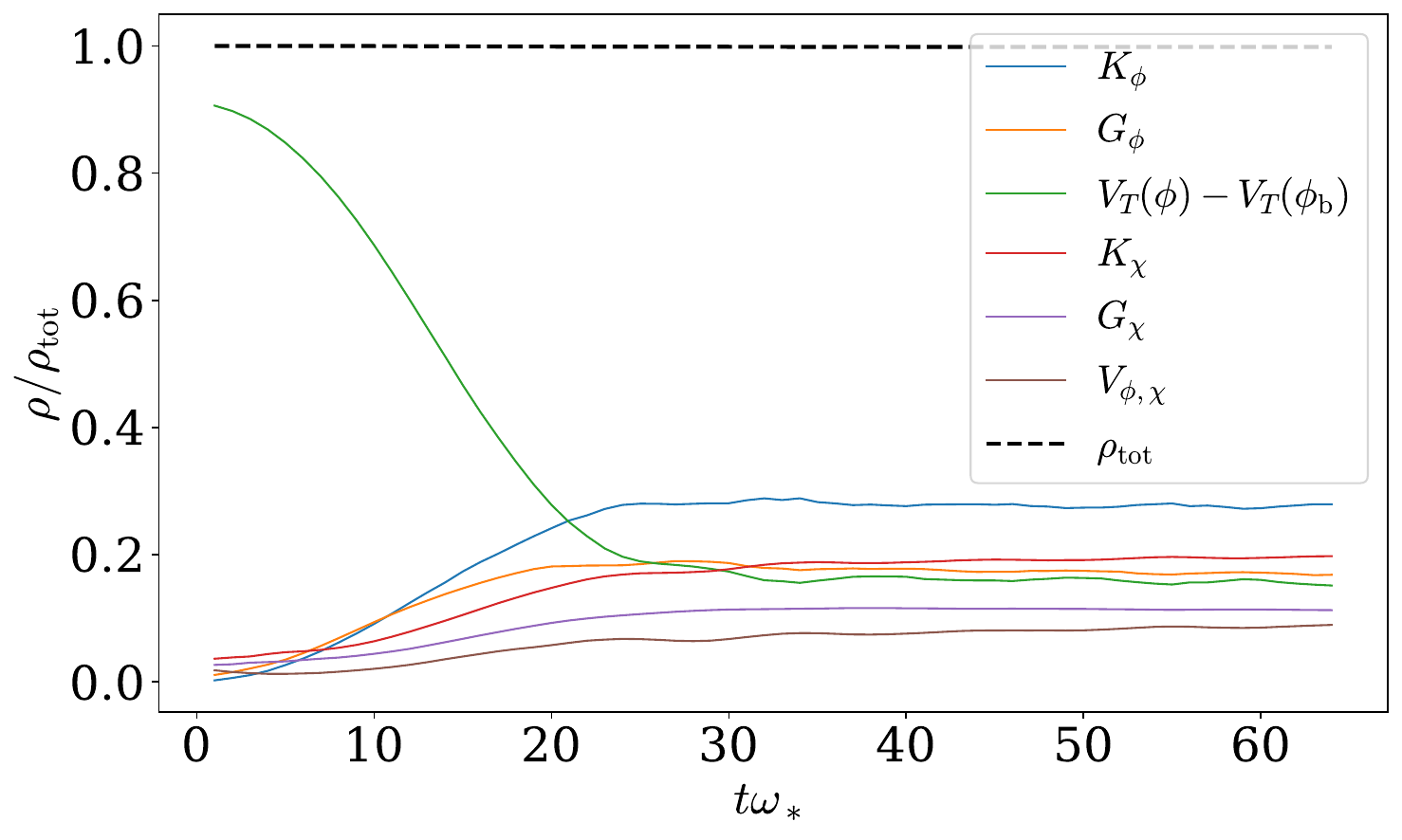}
    \caption{This figure shows the evolution of the energy fraction of each components, normalized to the total energy density $\rho_{\text{tot}}$, with the parameters $\lambda_{\phi\chi} = 5\lambda_\phi$ and $A = 1$. The components include kinetic ($K$) and gradient ($G$) energies of $\phi$ and $\chi$, interaction energy $V_{\phi\chi}$, and the shifted potential energy of $\phi$, $V(\phi) - V(\phi_\mathrm{b})$. The black dashed line represents the sum of these components, which remains constant throughout the simulation, thereby verifying the numerical accuracy.}
    \label{fig:energy_evolution}
\end{figure}

We also plot the time evolution of the fractional contributions from various energy components, as shown in Fig.~\ref{fig:energy_evolution}. During the FOPT, the vacuum energy is converted into the kinetic and gradient energies of $\phi$ and $\chi$, as well as the interaction potential energy. After the transition completes, the fractions of each energy component remain approximately constant. Note that in Fig.~\ref{fig:energy_evolution} the potential energy of $\phi$ and the total energy density $\rho_{\text{tot}}$ are shifted by subtracting the ground energy $V(\phi_\mathrm{b})$.

\section{Gravitational Waves}\label{sec:gw}
The GW production is quantified by solving the equation of motion for the transverse-traceless metric perturbations \( h_{ij}(\mathbf{x}, t) \)
\begin{align}\label{eq:GW_equation}
    \Box h_{ij} = \frac{2}{m_{\mathrm{pl}}^2} T_{ij}^{\mathrm{TT}},
\end{align}
where \( T_{ij}^{\mathrm{TT}} \) denotes the transverse-tracelesss part of the energy-momentum tensor of the scalar fields, and \( m_{\mathrm{pl}} = (8\pi G)^{-1/2} \) is the reduced Planck mass.

We define the dimensionless GW energy spectrum as~\cite{Caprini:2018mtu}
\begin{align}\label{spctral}
    \Omega_{\mathrm{GW}}(k) = \frac{m_{\mathrm{pl}}^2\, k^3}{8\pi^2\, \rho_c} \int \frac{d\Omega_k}{4\pi} \left| \dot{h}_{ij}(\mathbf{k}) \right|^2,
\end{align}
where \( \rho_c \) is the criticle energy density, and \( \mathrm{d}\Omega_k \) represents infinitesimal solid angle in the momentum space.

We first briefly review the analytical results derived from the envelope approximation, then compare with our numerically fitted results.
In the envelope approximation\cite{Kosowsky:1992vn}, bubbles expand as infinitesimally thin walls carrying all the released vacuum energy. Overlapping regions are neglected, and GWs are computed from the shear stress of the uncollided ``envelope".

In the envelope approximation, the peak amplitude can be estimated as
\begin{align}\label{eq:omega_env}
\Omega_\text{p}^{\text{env}} \simeq  \frac{0.44\, v_*}{(8\pi)^{2/3} (1 + 8.28\, v_*^3)}\left( R_* H_* \Omega_{\text{vac}} \right)^2,
\end{align}
where $v_*$ is the expected bubble wall velocity at collision, $H_*$ is the Hubble parameter when the FOPT occurs, and $\Omega_{\text{vac}}=\rho_{\text{vac}}/\rho_c$ is the fraction of vacuum energy. Since GWs are generated during bubble collisions, the peak frequency corresponds to the characteristic scale of the bubbles at collision, approximately $k_{\text{p}} \sim 1/R_*$. A more precise estimate is given by
\begin{align}
k_{\mathrm{p}}^{\mathrm{env}} \simeq \frac{1.96\,(8 \pi)^{1/3} v_*}{1 - 0.051 v_* + 0.88 v_*}\, R_*^{-1}.
\end{align}

\begin{figure*}[htbp]
\centering
\begin{minipage}{0.48\textwidth}
    \centering
    \includegraphics[width=\linewidth]{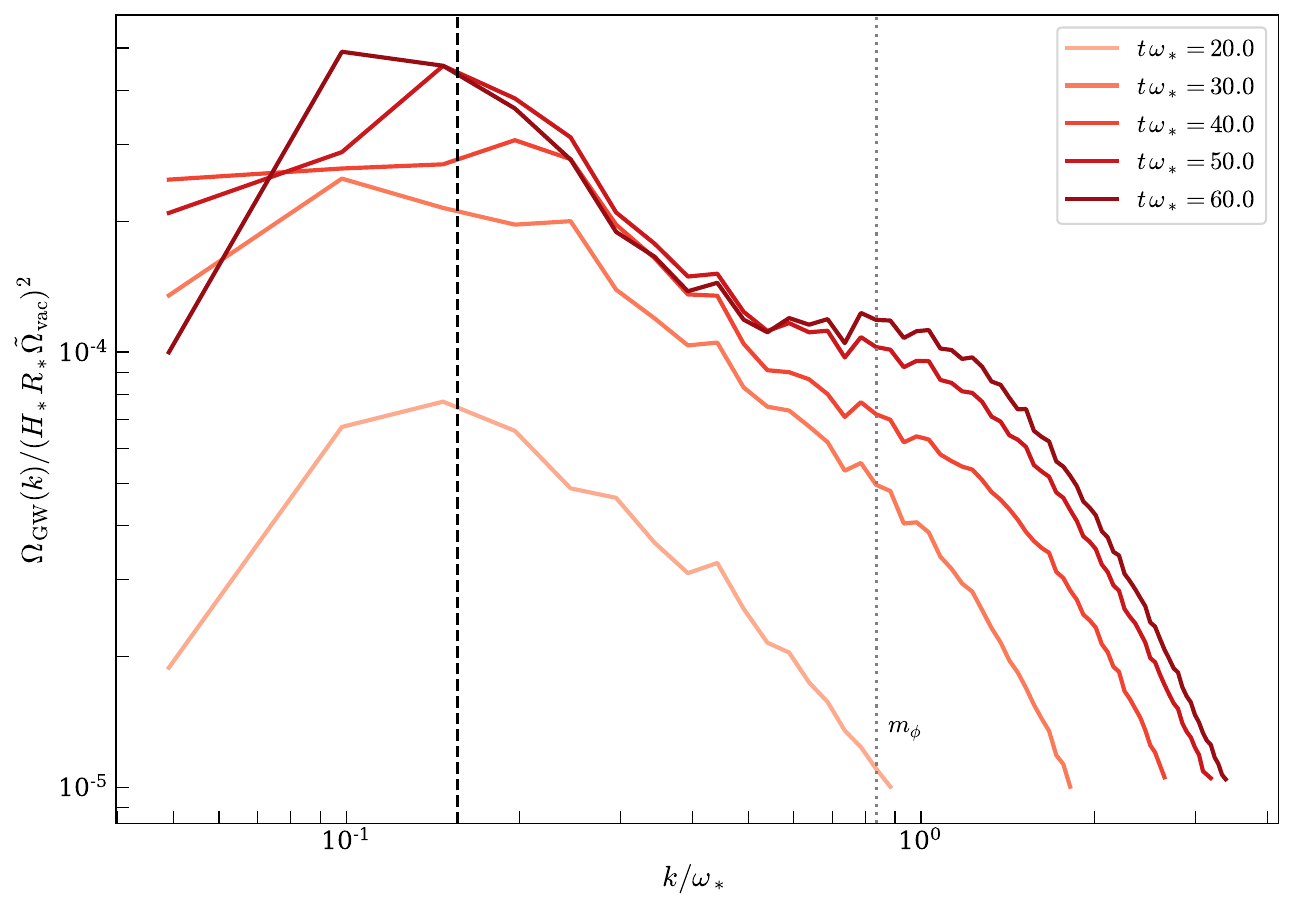}
    (a) The evolution of the GW energy spectrum in the uncoupled case.
    \quad
\end{minipage}
\begin{minipage}{0.48\textwidth}
    \centering
    \includegraphics[width=\linewidth]{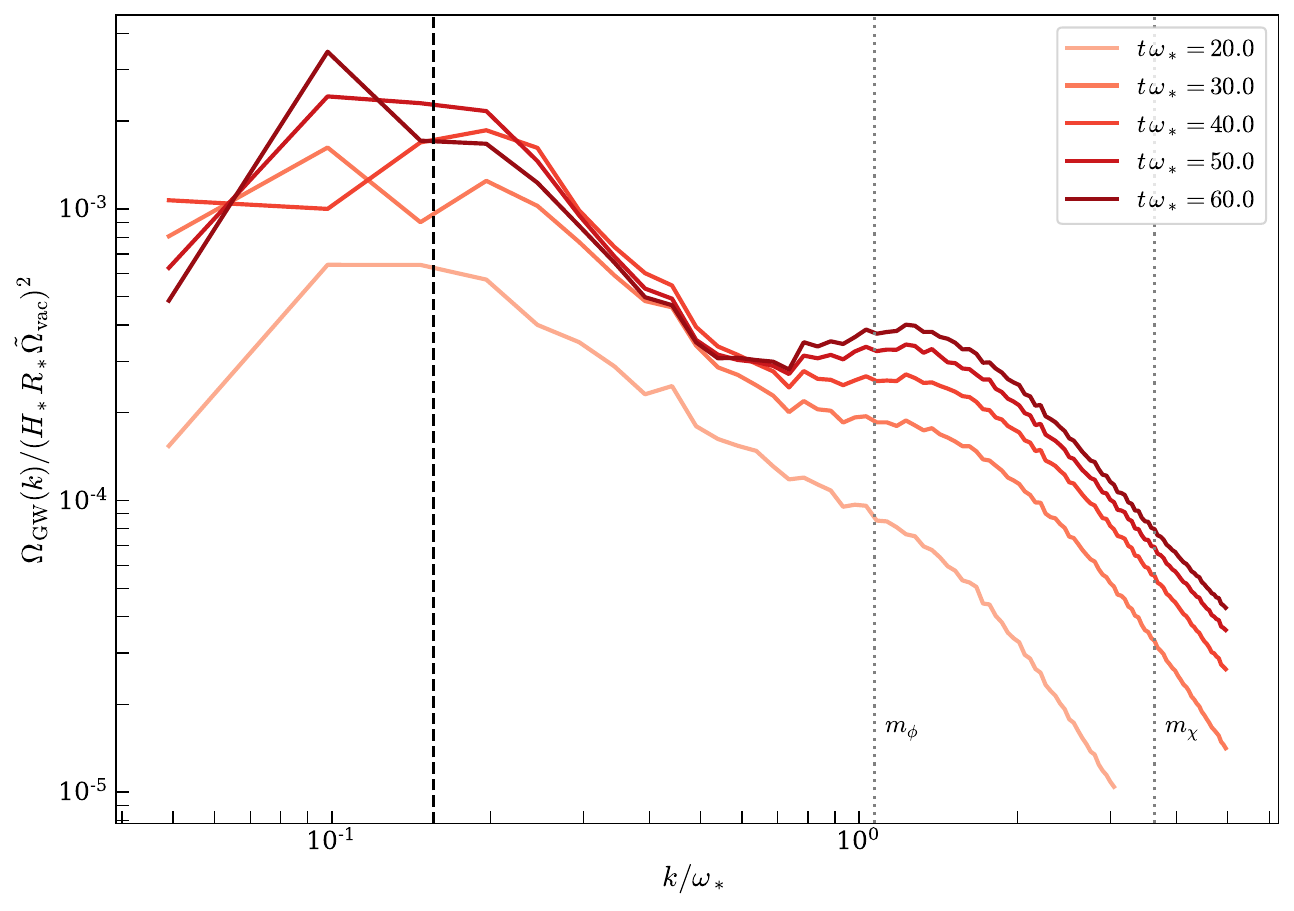}
    (b) The evolution of the GW energy spectrum in the coupled case with parameters ($\lambda_{\phi\chi}=5 \,\lambda_{\phi}$, $A=1$).
\end{minipage}
        \caption{The dimensionless GW energy spectrum $\Omega_{\mathrm{GW}}(k)$ is shown at different times $t \omega_* = 20, 30, 40, 50, 60$. The black dashed vertical lines correspond to the scale of the average bubble separation $R_*$, while the dotted lines mark the characteristic masses $m_\phi$ and $m_\chi$. In both panels, the peak wavelength is close to the scale of bubble collisions and gradually stabilize after $t \omega_* \gtrsim 40$, indicating the end of the active GW production at the peak. 
}
        \label{fig:gw_avg_multicenters}
\end{figure*}

In Fig.~\ref{fig:gw_avg_multicenters}, we present the time evolution of the GW energy spectra for both the uncoupled case and the coupled case with $\lambda_{\phi\chi} = 5\lambda_\phi$ and $A = 1$. Here, the dimensionless GW energy spectrum is normalized by $(H_* R_* \tilde{\Omega}_{\text{vac}})^2$, where $\tilde{\Omega}_{\text{vac}} = \tilde{\rho}_{\text{vac}} / \rho_c$ denotes the reduced fraction of vacuum energy. During the FOPT, a prominent peak of $\Omega_{\mathrm{GW}}$ emerges near the frequency $k=2\pi/R_*$, generated by bubble collisions. The amplitude of the primary peak generally stabilizes as bubble collisions complete. Meanwhile, persistent oscillations of the scalar fields generate gradually growing secondary peaks at frequencies associated with their effective masses, defined as
$$
m_{\phi,\chi} = \left. \sqrt{\partial_{\phi,\chi}^2 V(\phi,\chi)} \right|_{\phi = \phi_{\text{b}},\, \chi = 0}.
$$
In particular, a clear secondary peak develops near the mass scale of $\phi$. Due to the smaller energy fraction carried by $\chi$, the peak associated with $m_\chi$ is comparatively less pronounced.

\begin{figure*}[htbp]
    \centering
    \begin{minipage}[b]{0.48\textwidth}
    \centering
        \includegraphics[width=\textwidth]{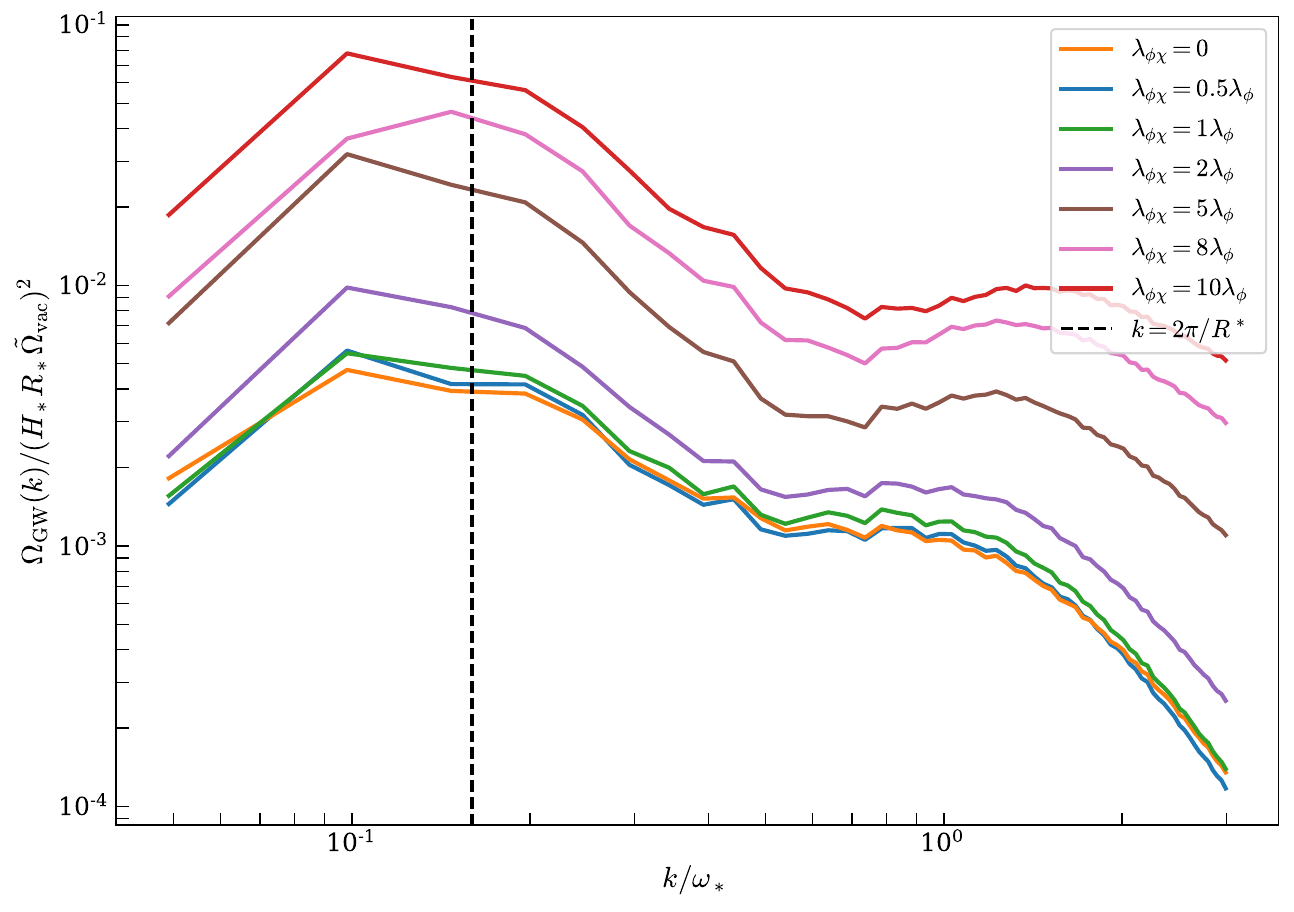}
        (a)The GW energy spectra for different $\lambda_{\phi\chi}/\lambda_\phi$ and fixed $A = 1$.
    \end{minipage}
    \begin{minipage}[b]{0.48\textwidth}
    \centering
        \includegraphics[width=\textwidth]{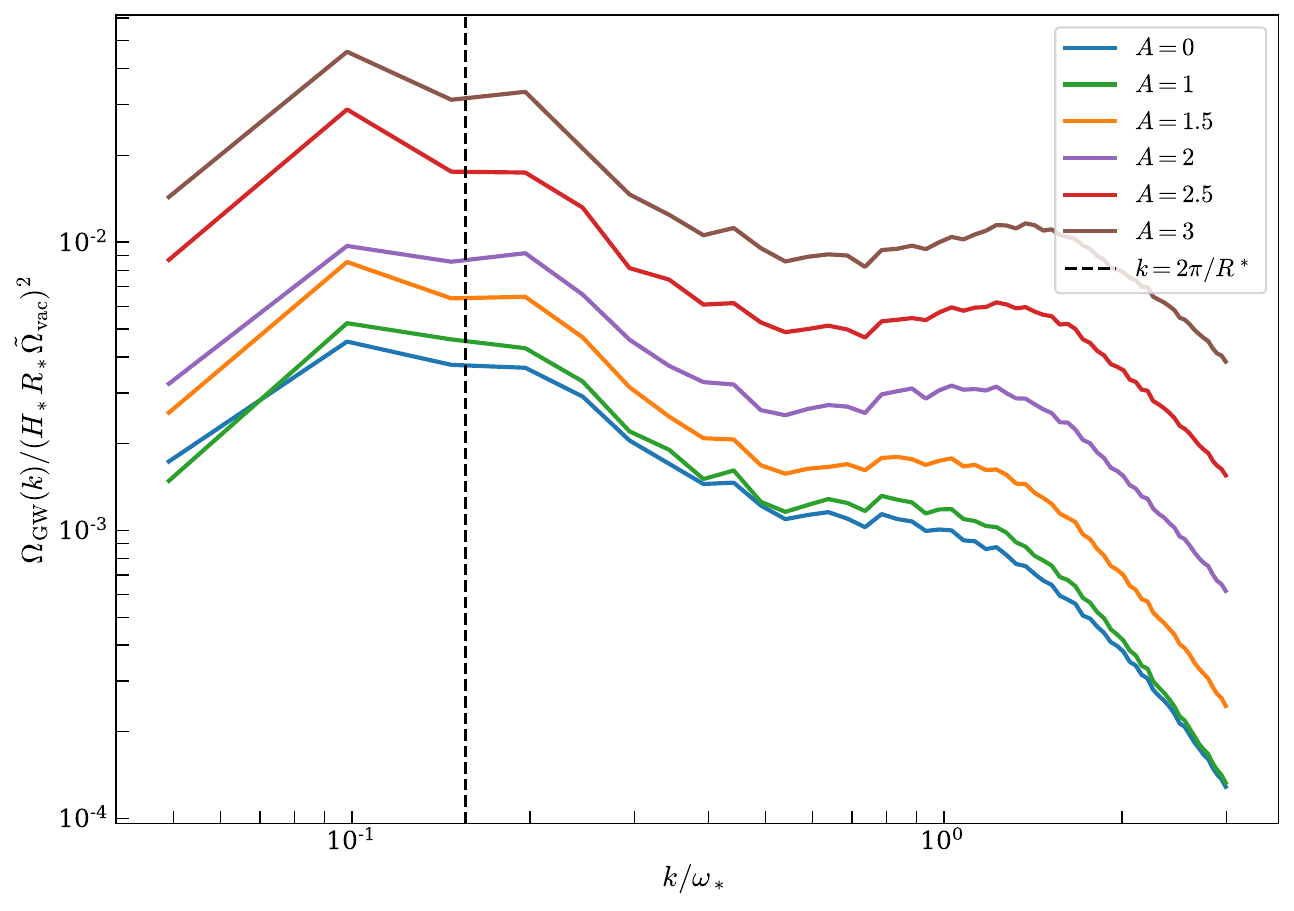}
        (b) The GW energy spectra for different fluctuation amplitude $A$ and fixed $\lambda_{\phi\chi} = \lambda_\phi$.
    \end{minipage}
    \caption{The dependence of $\Omega_{\mathrm{GW}}$ on coupling constant $\lambda_{\phi\chi}$ and the fluctuation amplitude $A$. These spectra are the averaged value over the time interval $t\omega_* = 44$ to $64$ for each set of parameters.
}
    \label{fig:gw_compare}
\end{figure*}

To suppress fluctuations arising from post-collision oscillations of the dominant peak, we average the GW spectra over the time interval $t\omega_* = 44$ to $64$ for each set of parameters to obtain the final spectrum, as shown in Fig.~\ref{fig:gw_compare}. 
The peak frequencies and the spectral slopes remain largely unaffected by changes in the parameters, whereas the peak amplitude increases with larger values of $\lambda_{\phi\chi}$ and $A$. This finding is expected since we have found the bubbles expand more rapidly with larger $A$ and $\lambda_{\phi\chi}$ in the single bubble evolution.

We fit the GW spectrum using the expression
\begin{align}
    \Omega^{\mathrm{fit}}_{\mathrm{GW}}(k) = \Omega_{\mathrm{p}} \frac{(a + b)^c\, k_{\mathrm{p}}^b k^a}{(b k_{\mathrm{p}}^{(a + b)/c} + a k^{(a + b)/c})^c},
\end{align}
where the parameter $a$ is fixed to 3 due to causality~\cite{Caprini:2009fx}. The fitting is performed over the momentum range $k < 0.42$ to ensure that the late-time peaks generated by field oscillations do not bias the results.
We obtain the best-fit parameters as
\begin{align}
b &= 1.3838 \pm 0.3032, \\
c &= 1.1007 \pm 0.3711, \\
k_{\text{p}}/\omega_* &= 1.108 \times 10^{-1} \pm 1.21 \times 10^{-2}\,.
\end{align}
Note that these fitting results are valid for both coupled and uncoupled case, since the interaction term has little affect on the peak frequency and spectral index of the GW energy spectrum. The peak amplitude in the uncoupled case is
\begin{align}
\tilde\Omega_{\mathrm{p}}/\left( R_* H_* \Omega_{\text{vac}} \right)^2 = 4.703 \times 10^{-3} \pm 1.97 \times 10^{-4}\,,
\end{align}
which is in good agreement with the prediction from the envelope approximation
\begin{align}
\frac{\tilde\Omega_{\text{p}}}{\Omega_{\text{p}}^{\text{env}}} &\simeq 0.851, \\
\frac{k_\text{p}}{k_{\text{p}}^{\text{env}}} &\simeq 1.390.
\end{align}

The fitted peak amplitudes $\Omega_\text{p}$ for different values of $\lambda_{\phi\chi}/\lambda_\phi$ and $A$ are summarized in Table~\ref{tab:fitted_peak_amplitudes}.
~\\

\begin{table}[htbp]
\centering
\begin{tabular}{c|c|c}
\hline
\hline
$\lambda_{\phi\chi}/\lambda_{\phi}$ & $A$ & $\Omega_\text{p}$ \\
\hline
0.0  & 1.00 & $4.703 \times 10^{-3}$ \\
0.5  & 1.00 & $5.565 \times 10^{-3}$ \\
1.0  & 1.00 & $5.622 \times 10^{-3}$ \\
1.0  & 1.50 & $8.801 \times 10^{-3}$ \\
1.0  & 2.00 & $1.046 \times 10^{-2}$ \\
1.0  & 2.50 & $3.192 \times 10^{-2}$ \\
1.0  & 3.00 & $2.735 \times 10^{-2}$ \\
2.0  & 1.00 & $9.008 \times 10^{-3}$ \\
5.0  & 1.00 & $2.203 \times 10^{-2}$ \\
8.0  & 1.00 & $4.693 \times 10^{-2}$ \\
10.0 & 1.00 & $7.843 \times 10^{-2}$ \\
\hline
\hline
\end{tabular}
\caption{Fitted peak amplitudes $\Omega_\text{p}$ for various values of $\lambda_{\phi\chi}/\lambda_\phi$ and $A$.}
\label{tab:fitted_peak_amplitudes}
\end{table}

Then, we derive the analytical interpretations of the GW enhancement with large interaction term. Quantum tunneling settle $\phi$ to the minimum of $V_{\text{re}}(\phi)$. Then $\chi$ acquires its effective mass and $\chi$'s oscillation amplitude decreases rapidly, which causes the true vacuum to descend towards the minimum of $V(\phi)$. Consequently, the actual released vacuum energy is close to $\rho_{\mathrm{vac}}$ rather than $\tilde{\rho}_{\mathrm{vac}}$. When $V_{\mathrm{re}}$ is fixed, larger values of $\lambda_{\phi\chi}$ or $A$ result in larger $\rho_{\mathrm{vac}}$ and release more vacuum energy during the FOPT, thereby leading to a stronger GW signal.

Intuitively speaking, the amplitude of GWs generated during the phase transition is expected to scale with the square of the released vacuum energy, as shown in the envelope approximation Eq.~\eqref{eq:omega_env}.

However, in the presence of coupling, part of the vacuum energy initially stored in $\phi$ is transferred to $\chi$. Consequently, the GW amplitude becomes proportional to $\left( \rho_{\text{vac}} - \Delta E_{\chi} \right)^2$, where $\Delta E_{\chi}$ denotes the energy transferred from $\phi$ to $\chi$. In other words,

\begin{align}\label{eq:ratio}
    \frac{\Omega_{\text{p}}}{\tilde{\Omega}_{\text{p}}} \simeq \left( \frac{\rho_{\text{vac}} - \Delta E_{\chi}}{\tilde{\rho}_{\text{vac}}} \right)^2,
\end{align}

where
\begin{widetext}
\begin{align}\label{eq:rho_vac}
\rho_{\text{vac}} = \frac{1}{192 \pi^4 \lambda_{\phi}^3} \bigg\{
&\left(
\sqrt{
\pi^2 \left( \beta^2 - 4 \lambda_{\phi} M_\phi^2 \right)
+ \lambda_{\phi} \lambda_{\phi\chi} A^2 k_{\text{max}}^2
} + \pi \beta
\right)^2 \nonumber\\
&\times \left[
2 \pi^2 \left( \beta^2 - 6 \lambda_{\phi} M_\phi^2 \right)
+ 2\pi \beta
\sqrt{
\pi^2 \left( \beta^2 - 4 \lambda_{\phi} M_\phi^2 \right)
+ \lambda_{\phi} \lambda_{\phi\chi} A^2 k_{\text{max}}^2
}
+ 3 \lambda_{\phi} \lambda_{\phi\chi} A^2 k_{\text{max}}^2
\right]
\bigg\}.
\end{align}
\end{widetext}

\begin{figure}[htbp]
    \centering
    \includegraphics[width=\linewidth]{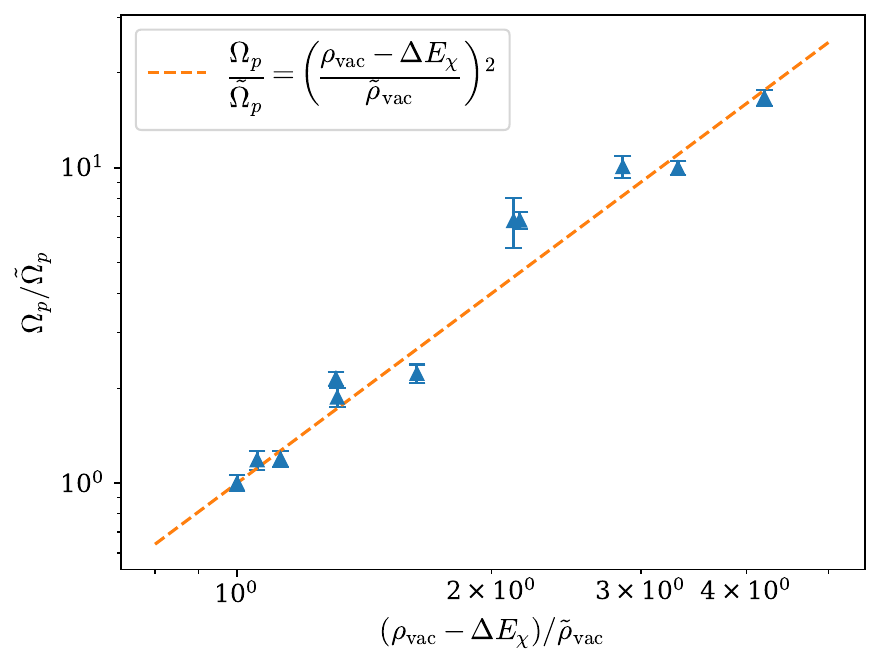}
    \caption{The scaling relation between the peak value of $\Omega_{\mathrm{GW}}$ and the effective vacuum energy release. The simulation data of $\Omega_p / \tilde{\Omega}_p$ (blue markers with error bars) are in good agreement to the theoretical prediction (orange dashed line) given by Eq.~\eqref{eq:ratio}.
}
    \label{fig:Omega_peak_scaling_errorbar}
\end{figure}

To test this interpretation, we plot $\Omega_{\text{p}} / \tilde{\Omega}_{\text{p}}$ as a function of $(\rho_{\text{vac}} - \Delta E_{\chi}) / \tilde{\rho}_{\text{vac}}$, using simulation results for $\Omega_{\text{p}}$, $\tilde{\Omega}_{\text{p}}$, and $\Delta E_{\chi}$, while computing $\rho_{\text{vac}}$ and $\tilde{\rho}_{\text{vac}}$ from Eq.~\eqref{eq:rho_vac}. 
The simulated results and their associated error bars are shown in Fig.~\ref{fig:Omega_peak_scaling_errorbar}. The theoretical relation predicted by Eq.~\eqref{eq:ratio} is also depicted. As seen in Fig.~\ref{fig:Omega_peak_scaling_errorbar}, the simulation results agree well with the theoretical prediction, supporting our interpretation.
Note that although the evolution of $\chi$ can generate GWs, its energy density is far less concentrated than that of bubble walls. This results in a significantly weaker energy-momentum tensor of $\chi$ so that its contribution of GWs can be neglected.

\begin{figure}[htbp]
    \centering
    \includegraphics[width=\linewidth]{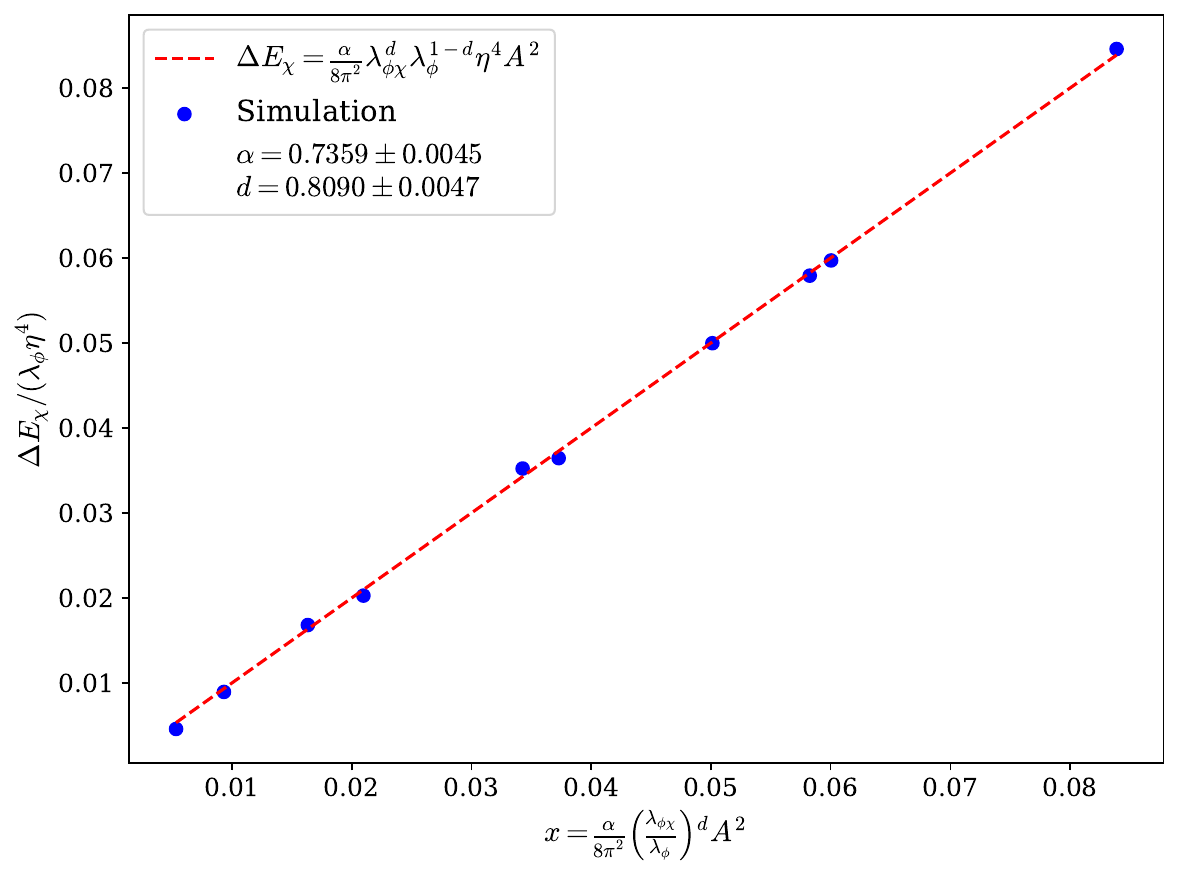}
    \caption{This figure shows that the theoretical prediction given by Eq.~\eqref{eq:delat_e} (red dashed line) excellently agrees with the simulation results for the transferred energy $\Delta E_\chi$ (blue dots) with parameters $\alpha = 0.7359 \pm 0.0045$ and $d = 0.8090 \pm 0.0047$.
    }
    \label{fig:Delta_e}
\end{figure}

Once the scaling relation for the peak amplitude, Eq.~\eqref{eq:ratio}, is established, we can provide a semi-analytical prediction for $\Omega_\text{p}$ by estimating $\Delta E_{\chi}$. Given the form of the interaction term in Eq.~\eqref{eq:full_effective_potential}, the energy transferred to $\chi$ is expected to depend on $A^2$, and $\lambda_{\phi\chi}$. Accordingly, we fit the simulation results for $\Delta E_{\chi}$ using the ansatz
\begin{align}\label{eq:delat_e}
\Delta E_{\chi} = \frac{\alpha}{8\pi^2} \lambda_{\phi\chi}^{d} \lambda_{\phi}^{1 - d} \eta^4 A^2,
\end{align}
where $\alpha$ and $d$ are the fitting parameters. The best-fit values are found to be
\begin{align}
\alpha &= 0.7359 \pm 0.0045, \\
d &= 0.8090 \pm 0.0047.
\end{align}
Substituting these into Eq.~\eqref{eq:delat_e}, we obtain the semi-analytical curve, which is plotted alongside the simulation data in Fig.~\ref{fig:Delta_e}. It shows that the theoretical prediction agrees well with the numerical results. Therefore, under the conditions considered in this work, $\Delta E_{\chi}$ can be estimated as
\begin{align}\label{eq:delta_e_result}
\Delta E_{\chi} = \frac{0.736}{8\pi^2} \lambda_{\phi\chi}^{0.81} \lambda_{\phi}^{0.19} \eta^4 A^2.
\end{align}
In this way, using the analytical expressions shown in Fig.~\ref{fig:Delta_e} and Fig.~\ref{fig:Omega_peak_scaling_errorbar}, we have successfully explained the numerical results of $\Omega_{\mathrm{GW}}$. In future works, this analytical result can also be directly used to predict $\Omega_{\mathrm{GW}}$ in our coupled case.

\section{Summary and Discussion}\label{sec:summary}
We have systematically investigated the dynamics and the GW production in scenarios where the scalar field $\phi$, responsible for the FOPT, is coupled to another massless scalar $\chi$ via the interaction term $\lambda_{\phi\chi} \phi^2 \chi^2$. The interaction term not only induces the effective mass of $\chi$ in the true vacuum bubbles, but also alters the effective potential $V_{\text{re}}(\phi)$ that governs the quantum tunneling. Under the condition that $V_{\text{re}}(\phi)$ is fixed, the numerical results implies that a larger interaction term, including both larger coupling constant $\lambda_{\phi\chi}$ and $\chi$'s oscillation amplitude $A$, leads to faster bubble expansion and amplified GW production. 

Through lattice simulations, we identified two dominant mechanisms by which the mass-acquiring $\chi$ modifies the transition:

\begin{enumerate}
  \item \textit{Energy transfer during the bubble expansion:}
  The $\chi$ field acquires a large effective mass in the true vacuum. Consequently, $\chi$-particles are reflected off the advancing bubble walls during expansion. This reflection mechanism transfers a fraction of the released vacuum energy to $\chi$ via the interaction term, effectively diminishing the energy available to generate GWs. This energy reduction is analogous to friction exerted by a background fluid.

  \item \textit{Releasing additional energy in the true vacuum}:  
  More importantly, we identify a novel effect that largely enhances the FOPT. The combined influence of $\chi$-particle reflection and its large effective mass rapidly suppresses the oscillation amplitude of $\chi$ within the true vacuum. This suppression lowers the minimum of the effective potential thereby rapidly releasing the energy stored in $V_\mathrm{re}$. The releasing of the additional energy in turn significantly intensify the FOPT, which outweighs the attenuating influence of the friction-like mechanism.


\end{enumerate}

We also establish an analytical formulism that accurately explains our results of the GW energy spectrum. These findings underscore that coupled fields with dynamically evolving mass can significantly backreact to the vacuum structure and field dynamics. This necessitates their inclusion in theoretical predictions of GW signals from early-Universe FOPTs. {This mechanism is generally relevant for scenarios involving mass-acquiring spectator fields, such as filtered dark matter and Q-ball formation. In such cases, analytical estimates of the gravitational-wave spectrum that rely solely on the initial potential may underestimate the signal strength. The lattice approach developed here provides a consistent framework to capture both the trivial potential modification and the additional dynamical energy transfer during the phase transition.
} 

{For scenarios involving larger couplings, one-loop quantum corrections to the effective potential may become non-negligible. These corrections introduce additional terms proportional to $\phi^4$, $\phi^2\chi^2$, and $\chi^4$~\cite{Coleman:1973jx,Quiros:1994dr}. The first two can be absorbed into the renormalized potential, while the $\chi^4$ term may affect the evolution of $\chi$ only when the coupling is strong and its fluctuations are large. A systematic investigation of these effects in the strong-coupling regime is left for future work.}


While our analysis provides a comprehensive account of the impact of dynamically mass-acquiring scalar fields, it remains restricted to scalar degrees of freedom. In particular, scenarios where the coupled field is fermionic—such as in models involving chiral fermions or heavy dark sector states—pose additional theoretical and computational challenges. Since classical lattice simulations are not directly applicable to fermionic fields due to the absence of a classical limit, such setups require alternative approaches, potentially involving effective fluid models or quantum lattice techniques\cite{Kasper:2014uaa,Kreula:2016qci,Zache:2018jbt}. Extending the present framework to incorporate spinor fields remains an important topic for future work.
\section*{ACKNOWLEDGMENTS}
This work is supported in part by the National Natural Science Foundation of China No. 12235019 and No. 12475067, in part by the National Natural Science Foundation of China No. 12147103, No. 12235019, No. 12075297 and No. 12347103.

\appendix

\section{Lattice Effect}\label{app:lattice_effect}

As the bubble expands, Lorentz contraction causes its wall to thin continuously. As the thickness of the bubble wall gradually approaches the resolution limit of the lattice simulation, this process will lead to a significant increase in the numerical error of the field evolution. Eventually, the thickness of the bubble wall can no longer decrease and is forced to stay at the order of the lattice spacing. At this point, the accelerated expansion of the bubble wall is halted, and the released vacuum energy is dissipated into smaller-scale oscillations, as noted in Refs.~\cite{combs1983single, Peyrard:1984kdi}.

\begin{figure}[htbp]
    \centering
    \includegraphics[width=\linewidth]{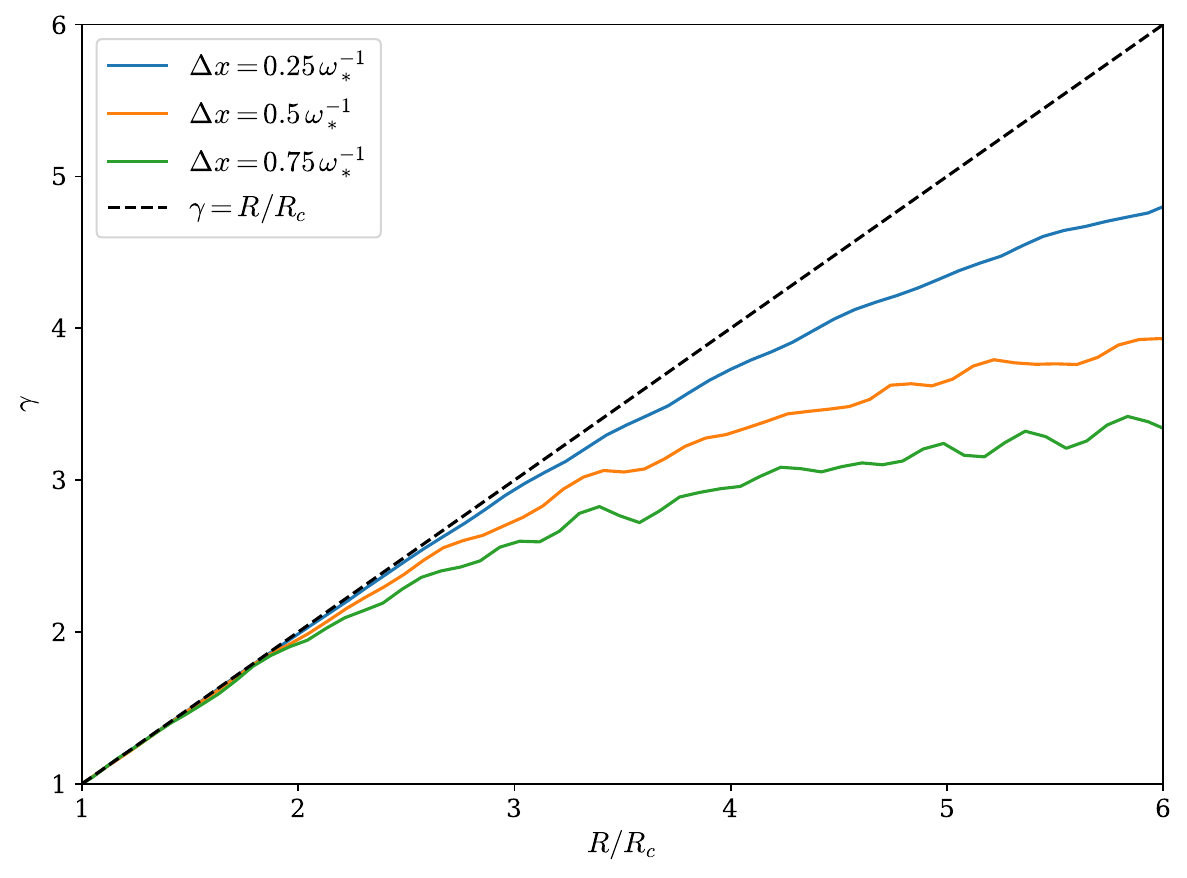}
    \caption{The dependence of bubble wall Lorentz factor $\gamma$ on bubble radius $R$ for different lattice spacings $\Delta x$. Solid curves represent simulation results for $\Delta x = 0.25\,\omega_*^{-1}$ (blue), $0.5\,\omega_*^{-1}$ (orange), and $0.75\,\omega_*^{-1}$ (green). The black dashed line indicates the theoretical prediction $\gamma = R / R_c$ from energy conservation. Smaller lattice spacings yield better agreement with theory, demonstrating improved numerical resolution and energy conservation.
}
    \label{fig:lattice_effect}
\end{figure}

To assess the impact of lattice resolution, we measure the expansion velocity of a single bubble in the absence of coupling by varying the lattice spacing among $\Delta x = 0.25\,\omega_*^{-1}$, $0.5\,\omega_*^{-1}$, and $0.75\,\omega_*^{-1}$. We compare the Lorentz factor $\gamma$ of the bubble wall as a function of the bubble radius $R$, with results shown in Fig.~\ref{fig:lattice_effect}.

Colored solid curves indicate the relation between $\gamma$ and $R$ for different lattice resolutions, while the black dashed line shows the theoretical prediction from energy conservation (Eq.~\ref{eq:gamma_r}). It evidently shows that , smaller lattice spacings lead to slower degradation of the bubble wall velocity and better agreement with the theoretical curve, indicating improved energy conservation. For the resolution $\Delta x = 0.25\,\omega_*^{-1}$ used in our simulations, lattice artifacts remain under control up to $\gamma \simeq 4$.

\bibliography{ref}
\end{document}